\documentclass[11pt,a4paper]{article}
\pdfoutput=1
\usepackage{jcappub_mod}
\usepackage[english]{babel}

\let\oldsqrt\sqrt
\def\sqrt{\mathpalette\DHLhksqrt}
\def\DHLhksqrt#1#2{%
\setbox0=\hbox{$#1\oldsqrt{#2\,}$}\dimen0=\ht0
\advance\dimen0-0.2\ht0
\setbox2=\hbox{\vrule height\ht0 depth -\dimen0}%
{\box0\lower0.4pt\box2}}

\graphicspath{{./figs/}}

\def\figwidth{0.93\columnwidth}

\begin{document}

\title{
\color{BlueViolet}
Scalar dark matter in an extra dimension inspired model
}

\author[a]{R.~A.~Lineros}
\author[a,b]{F.~A.~Pereira dos Santos}
\affiliation[a]{
Instituto de F\'{\i}sica Corpuscular -- CSIC/U. Valencia,\\ Parc Cient\'{\i}fic, calle 
Catedr\'{a}tico Jos\'{e} Beltr\'{a}n 2, E-46980 Paterna, Spain\\}
\affiliation[b]{
Departamento de F\'isica, Pontif\'icia Universidade Cat\'olica do Rio de Janeiro,\\
Rua Marqu\^es de S\~ao Vicente 225, 22451-900 G\'avea, Rio de Janeiro, Brazil\\}

\abstract{
%The nature of the dark matter of the Universe posses an intriguing problem in cosmology and particle physics.
%
In this paper we analyze a dark matter model inspired by theories with extra dimensions. 
The dark matter candidate corresponds to the first Kaluza-Klein mode of an real scalar added to the Standard Model.
The tower of new particles enriches the calculation of the relic abundance.
For large mass splitting, the model converges to the predictions of the inert singlet dark matter model.
For nearly degenerate mass spectrum, coannihilations increase the cross-sections used for direct and indirect dark matter searches.
Moreover, the Kaluza-Klein zero mode can mix with the SM higgs and further constraints can be applied.
}

%\arxivnumber{1407.xxxx}
%\dedicated{IFIC/14-yyy}

\maketitle

\section{Introduction}

Despite of observational and theoretical efforts the nature of dark matter remains unknown. Its existence, if confirmed, establishes the need, besides neutrino mass, for physics beyond the standard model. 
However, until now we have only hints for its existence. Many extensions of the standard model of particle physics provide new candidates that are consistent with dark matter. Some of these models arise from theories formulated to address others problem in particle physics, for instance the hierarchy problem and the strong CP problem, and more recently neutrino mass.\\

A very popular class of dark matter candidates stands for weakly interacting massive particles (WIMPs)~\cite{Bertone:2004pz}. Along this line of reasoning, a widely studied WIMP is the neutralino that comes from the minimal supersymmetric standard model (MSSM)~\cite{Jungman:1995df}.
The neutralino dark matter is stable due to R-parity conservation and constitutes a typical and very well studied WIMP candidate. Also, the solution of the strong CP problem via Peccei-Quinn mechanism~\cite{Peccei:1977hh} requires a new dynamical scalar field with appropriate charges that
gives rise to axion as a new particle~\cite{Peccei:1977ur, Weinberg:1977ma, Wilczek:1977pj}.
The axion is a pseudo-scalar particle that can be a dark matter candidate due to non-thermal production in the early universe~\cite{Dine:1982ah, Abbott:1982af, Preskill:1982cy}.\\

In the context of beyond the standard model physics, extra dimensional theories appeared as an alternative to other more popular scenarios aforementioned. In recent years, extra dimension 
models have been proposed to solve several problems in particle physics like the hierarchy problem~\cite{ArkaniHamed:1998rs, Antoniadis:1998ig, Randall:1999vf, Randall:1999ee}. More recently, many realizations of models 
with extra dimensions have been applied to dark matter problem~\cite{Servant:2002aq, Cheng:2002ej} (see also~\cite{Ponton:2008zv, Medina:2011qc}). 
Among many dark matter extra dimensional models, the most usual realization  is based on the 
universal extra dimension (UED) in which all the standard model particles propagate in a flat extra dimensional space~\cite{Appelquist:2000nn}. In these models, interactions in the bulk do not violate momentum 
conservation along the extra dimension which implies the conservation of Kaluza-Klein (KK) number at tree level. Also, all of the couplings among the standard model particles arise 
from bulk interactions. The stability of the lightest Kaluza-Klein particle (LKP) is a result of a remnant
 discrete symmetry called KK parity that arise after breaking 
 translational invariance due to fixed points in the orbifold compactifications~\cite{Georgi:2000ks}. Since KK parity conservation ensures that the lightest Kaluza-Klein particle 
is stable it can be a suitable WIMP candidate~\cite{Servant:2002aq, Cheng:2002ej}. As Kaluza-Klein particles are naturally degenerate around a mass scale set by inverse of the compactification scale radius, the effect of coannihilations become important. As shown in~\cite{Servant:2002aq, Burnell:2005hm} 
the effect of coannihilations with KK states is to provide a small annihilation cross section that can be translated to an increase in the relic abundance of LKP.\\

In this work we propose a model based on flat extra dimension that is different from the UED model in the field propagation. We assume that all the standard model particles are confined to a 3-brane and only a singlet scalar \cite{Deshpande:1977rw, Silveira:1985rk, McDonald:1993ex, Barger:2008jx, Burgess:2000yq} (and gravity) can propagate in the whole space. We would like to stress that unlike the standard model (SM) Higgs doublet, that couples to fermion and gauge field, the scalar singlet that we consider in this work is inert in the sense that it 
 interacts only with the Higgs field via the scalar potential and it does not acquire a vacuum expectation value as we show in the section~\ref{model}.
This is a phenomenological setup aimed to address the effects of a single KK tower. 
The stability of the first nonzero state of the KK tower associated to the singlet scalar is ensured by a remnant U$(1)$-like symmetry that arises after integrating over the extra dimension.  The first nonzero KK particle  will be our dark matter particle.
We investigate the impact of the bulk singlet scalar for indirect and direct detection of dark matter.We calculate the annihilation cross section and spin independent cross section including coannihilation as a function of dark matter mass. Also, we show that annihilation and spin independent cross sections of dark matter undergo an increase when we include coannihilation. We include in our results for annihilation cross section  bounds from FERMI-LAT~\cite{2013arXiv1310.0828T}, MAGIC~\cite{2013arXiv1312.1535A} and HESS~\cite{2011PhRvL.106p1301A} and for spin independent cross section bounds from XENON100~\cite{XENON:2012Ph}, LUX~\cite{2013arXiv1310.8214L}, CDMSlite~\cite{2013arXiv1309.3259A}, CRESST~\cite{2014arXiv1407.3146C}, and DAMA/LIBRA~\cite{2009JCAP...04..010S}.\\

The paper is organized as follows. In the section~\ref{LED} we introduce the flat extra dimension framework that we are considering in this paper. Section~\ref{model} is dedicated to the description of the singlet scalar in the extra dimension formalism. In section~\ref{relic} we discuss the coannihilations formalism that is relevant for calculations of dark matter relic density. Our results are presented in the section~\ref{results}.

\section{Extra dimensions in a nutshell}
\label{LED} 

\subsection{General aspects of extra dimensions}

One of the main motivations for the interest in extra dimensional models comes from the 
possibility to address in an alternative way the hierarchy problem of the standard model 
of particle physics \cite{ArkaniHamed:1998rs, Randall:1999vf, Randall:1999ee}. However, in the course of their development extra dimensional theories
have shown that they can address others problems in particle physics such as symmetry breaking \cite{Cheng:1999bg}, 
mass and mixing of fermions \cite{Biggio:2003kp}, and they can provide a natural candidate to dark matter \cite{Appelquist:2000nn}.
Also they may modify the evolution of the Universe~\cite{1999PhRvL..83.4245C}.
\\

The modern theory of extra dimensions are based on the concept of the braneworld where we can think 
that our (3+1)-dimensional world is associated to a 3-dimensional membrane or simply 3-brane.  
 Extra dimensional models can be classified in two different groups that depend of 
the geometry.
The first group is called factorizable geometry or flat extra dimension. Inside this group we
have three main models that differ in the fields propagation. The first one is called large
extra dimension (LED) \cite{ArkaniHamed:1998rs} where only gravity propagates in the whole space and the 
standard model 
fields are confined to a 4-dimensional spacetime. In this model the size of 
compactified dimension can be large as $10^{-3}$ m. In the TeV model\footnote{For a review on TeV extra dimensions we suggest \cite{Ponton:2012bi}.} \cite{Antoniadis:1990ew, ArkaniHamed:1998nn, Dienes:1998vg} the standard
 model gauge 
bosons can propagate in the extra dimension and the size of the extra dimensions are of the 
order O$(10^{-19})$ m or TeV$^{-1}$, and the standard model fermions also are trapped in the 3-brane. 
In the universal extra dimensions (UED) model \cite{Appelquist:2000nn} all the fields are allowed to propagate in the whole space and 
the fundamental or zero modes of the Kaluza-Klein tower are associated with the standard model
particles.\\

The second group we will mention is based on a non-factorizable geometry or warped extra 
dimension (WED) and its variants \cite{Randall:1999vf, Randall:1999ee}. As in the factorizable models the WED scenarios can 
provide a plethora of new models that differ in the field propagation. In this work
we are interested in a modified version of the flat large extra dimensions that we will discuss
in the following.

\subsection{The modified large extra dimension framework} 

As mentioned above, in theories with large extra dimensions (LED) it is assumed that
the standard model fields are confined to a $4$-dimensional brane,
while the gravitational field propagates in the $4+\alpha$ space-time,
where $\alpha$ denotes the extra flat spatial dimensions. In order to solve
the hierarchy problem the fundamental scale is generally considered to be
close to the electroweak scale and is related with Planck scale by
\begin{eqnarray}
M^{2}_{Pl(4)}=M^{2+\alpha}_{FS}V^{\alpha},
\end{eqnarray}
where $M_{Pl(4)}$ is the 4D Planck scale, $M_{FS}$ is the fundamental scale in the LED model and $V^{\alpha}$ is the volume of extra spatial dimensions.

One important aspect of extra dimensional models are their dependence on the geometry of the
inner space that is reflected in turn in the kind of compactification of the extra dimension.
In order to clarify our point, we consider one field propagating in one extra dimension and we 
assume that the extra 
dimension is compactified on a circle of radius $R$. As a consequence the momentum in the compact
dimension is quantized and is given by
\begin{eqnarray}\label{p5}
p_5 = \frac{n}{R},
\end{eqnarray}
where $n$ is the Kaluza-Klein number which becomes a quantum number under a U$(1)$ symmetry 
from the 4-dimensional perspective . The quantization of the extra dimensional momentum, 
implies that the momentum of the particle must be larger than $R^{-1}$ that is the energy
 threshold for the observation of the Kaluza-Klein modes. It is important to mention that  
equation \ref{p5} means that the particle 
represented by 
the plane wave function can be seen from the 4-dimensional point of view as tower of $n$ 
particles with mass difference given by

\begin{eqnarray}
\Delta m \sim \frac{1}{R}.
\end{eqnarray}

One important aspect of the compactification is related to the presence of 
boundary localized terms (BLT). These terms arise as a consequence of the boundary conditions
imposed on the geometry of the inner space and also could be generated by radiative 
corrections. In this perspective the effect of BLT is loop suppressed and become important 
at loop level \cite{Georgi:2000ks}. In general, it is consider that these terms vanish at the
cut-off scale $\Lambda$ that defines the minimal extra dimensional model. \\
 
However, for the orbifold compactification these terms are important and are 
localized in the two fixed points. The main consequence of including these terms is that they 
break the 
translational invariance and hence the momentum in five dimensions is not conserved anymore.
Also these terms modify the Kaluza-Klein spectrum as well as the interactions among the 
particles. This kind of compactification is important to obtain chiral fermions in 
4-dimensions from a 5-dimensional theory \cite{Georgi:2000wb}\footnote{For a recent discussion on the boundary localized terms from the collider and dark matter perspective  in a different context of extra dimension we suggest~\cite{Datta:2012xy, Datta:2013nua}.}.\\

For our purpose we consider one single extra dimension and also we assume that one singlet 
scalar field can propagate in the whole (4+1)-spacetime.
 We would like to stress that the main point of our 
work is to provide a dark matter 
candidate in a natural way without invoking any symmetry to stabilize the particle. 
We will see that 
the stabilization condition is fulfilled  after integrating over the extra dimension. It seems 
that after compactification a U(1) symmetry remains to stabilize the first Kaluza-Klein
particle that is our dark matter candidate.

\section{The model}
\label{model}

The model is composed of the SM particle content enlarged with a SM singlet real scalar, $S$, which can propagate through the extra dimensions.
We assume that $S$ does not have any {\it apriori} symmetry i.e. we consider the most general scalar potential that contains all possible terms allowed by the SM symmetries.
The compactification of the extra dimension will generate the KK tower of $S$ and their interactions described by an effective lagrangian in (3+1) dimensions. 

\subsection{The scalar sector in (4+1) dimensions}
\label{ssec:scalar}

The scalar sector is composed by the real scalar $S$, which is the only field in the model that depends on the extra dimension $y$, and the SM higgs doublet $H$. From now on, the dependency on the usual (3+1)-dimensions is implicitly assumed. The langrangian of the scalar sector is divided into three parts:
\begin{eqnarray}
	\mathcal{L}_{\rm scalar}(y) &=& \mathcal{L}_{H} + \mathcal{L}_{S}(y)  + \mathcal{L}_{HS}(y) \ ,
\end{eqnarray}
where the terms stand for the lagrangian of the Higgs, the real singlet $S(y)$, and their interactions, respectively. \\

The lagrangian for $H$ is:
\begin{eqnarray}
\mathcal{L}_{H} &\propto& (D_{\mu} H)^{\dagger} D^{\mu} H - \frac{m_D^2}{2 v^2} \left(H^{\dagger}H - \frac{v^2}{2}\right)^2 \ ,
\end{eqnarray}
where in the unitary gauge: $H^{T} = (0 , (v + h))/\sqrt{2}$. The SM higgs vacuum expectation value is $v \simeq 246$~GeV and $m_D$ the mass term.\\

The lagrangian for $S$ is:
\begin{eqnarray}
\label{eq:lagscal}
	\mathcal{L}_{S}(y) = \frac{1}{2} \partial_{N} S(y) \partial^{N} S(y) - \frac{m_{S}^2}{2} S(y)^2 - \mu S - \frac{\lambda_{3}}{3} S(y)^3 - \frac{\lambda_4}{4} S(y)^4 \ ,
\end{eqnarray}

where $N$ runs over all (4+1) dimensions. The interaction lagrangian is:
\begin{eqnarray}
	\mathcal{L}_{HS}(y) = - \lambda_{H}\left(H^{\dagger}H - \frac{v^2}{2}\right)S(y) - \lambda_{2H} \left(H^{\dagger}H - \frac{v^2}{2}\right)S(y)^2 \ ,
\end{eqnarray}

For simplicity, we impose that $S$ will not acquire vev. This condition is ensured with following relations:
\begin{equation}
\mu = 0, \ \lambda_4 > 0, \ m_S^2 > 0 \  {\rm and} \ \lambda_3^2 < 4 \, \lambda_4 \, m_S^2 \, ,
\end{equation}

allowing the electroweak symmetry breaking to occur in the same way as in the SM. Also we include the condition: $\lambda_4 < 3$ to stay in the non-perturbative limit of the model.\\

\subsection{The effective langrangian}
The effective model in (3+1)-dimensions is obtained by integrating over the extra dimension.
The extra dimension is compact and described by a ring of radius $R$.
Assuming periodic boundary condition: $S(y) = S(y + 2 \pi R)$, we can expand $S(y)$ in fourier modes:
\begin{equation}
S(y) = \frac{1}{\sqrt{\pi R M_5}} \left\{ \frac{S_0}{\sqrt{2}} + \sum_{n=1}^{\infty} S_n \cos{\left(\frac{n y}{R}\right)} + \sum_{n=1}^{\infty} P_n \sin{\left(\frac{n y}{R}\right)} \right\} \, ,
\end{equation}
where $S_n$ and $P_n$ correspond to the KK modes of $S(y)$.
The term $M_5$ is a dimensional parameter used to preserve the correct dimensionality.\\

The effective lagrangian is simply obtained by:
\begin{eqnarray}
\mathcal{L} = \mathcal{L}_{\rm SM} + M_5 \int_{0}^{2\pi R} dy \left(\mathcal{L}_{S}(y) + \mathcal{L}_{HS}(y)\right) \, ,
\end{eqnarray}
where the effective lagrangian for $\mathcal{L}_{S}(y)$ is then:
\begin{eqnarray}
\label{eq:lseff1}
\mathcal{L}_S^{\rm eff} &=& \frac{1}{2}\left( \partial_{\mu} S_0 \partial^{\mu} S_0 - m_S^2 S_0^2 \right) + \frac{1}{2}\left( \sum_{n=1}^{\infty} \partial_{\mu} S_n \partial^{\mu} S_n - (m_S^2 + \frac{n^2}{R^2}) S_n^2 \right) \nonumber \\
& & + \frac{1}{2}\left( \sum_{n=1}^{\infty} \partial_{\mu} P_n \partial^{\mu} P_n - (m_S^2 + \frac{n^2}{R^2}) P_n^2 \right) - \frac{\omega_3}{3} \mathcal{V}_3 - \frac{\omega_4}{4} \mathcal{V}_4 \, . 
\end{eqnarray}
The KK modes, $S_n$ and $P_n$, are degenerate in mass because they receive the same contribution from compactification:
\begin{equation}
m_{S_n}^2 = m_{P_n}^2 = m_S^2 + \frac{n^2}{R^2} = m_S^2 + n^2 M_{KK}^2 \, ,
\end{equation}
where $M_{KK} = R^{-1}$.
Therefore, this allows us to describe them as a single complex scalar field: 
\begin{equation}
\chi_n = \frac{S_n + i P_n}{\sqrt{2}} \  {\rm and} \ \chi_n^{*} = \frac{S_n - i P_n}{\sqrt{2}} \, ,
\end{equation}
reducing equation~\ref{eq:lseff1} to:
\begin{equation}
\mathcal{L}_S^{\rm eff} = \frac{1}{2}\left( \partial_{\mu} S_0 \partial^{\mu} S_0 - m_S^2 S_0^2 \right) + \sum_{n=1}^{\infty} \partial_{\mu} \chi_n^{*} \partial^{\mu} \chi_n - m_{S_n}^2 \chi_n^{*}\chi_n - \frac{\omega_3}{3} \mathcal{V}_3 - \frac{\omega_4}{4} \mathcal{V}_4 \, .
\end{equation}
The couplings $\omega_i$ are related with $\lambda_i$ via factors that come from the compatification. In this case:
\begin{equation*}
\omega_3 = \frac{\lambda_3}{\sqrt{2 \pi R M_5}} \ {\rm and} \ \omega_4 = \frac{\lambda_4}{2 \pi R M_5} \, .
\end{equation*}
The terms $\mathcal{V}_3$ and $\mathcal{V}_4$ come from the interaction terms of 3 and 4 fields (equation~\ref{eq:lagscal}).
For example, considering up to $n=2$ KK modes, $\mathcal{V}_3$ is: 
\begin{eqnarray}
\mathcal{V}_3 &=& S_0^3 + 3 S_0 \left(|\chi_1|^{2} + |\chi_2|^{2}\right)  + 3 \left(\chi_1^2 \chi_2^{*} + \chi_2 {\chi_1^{*}}^2  \right) \, ,
\end{eqnarray}
and $\mathcal{V}_4$:
\begin{eqnarray}
\mathcal{V}_4 &=& S_0^4 + 12 S_0^2 \left(|\chi_1|^2 + |\chi_2|^2\right) + 12 S_0 \left(\chi_1^2 \chi_2^{*} + \chi_2 {\chi_1^{*}}^2\right) \nonumber \\ 
&&+ 6 \left(|\chi_1|^4 + 4 |\chi_1|^2 |\chi_2|^2 + |\chi_2|^4\right) \, ,
\end{eqnarray}
but $\mathcal{V}_{3,4}$ can be calculated for an arbitary number of KK modes.
We observe that a global $U(1)$ symmetry with charges $Q(\chi_n) = n$ is present in the lagrangian.
The lightest KK state, $\chi_1$, is stable and therefore it is the dark matter candidate.\\

The effective lagrangian for the interactions with the higgs field is:
\begin{equation}
\mathcal{L}_{HS}^{\rm eff} = - \left(H^{\dagger}H - \frac{v^2}{2}\right) \left( \omega_H S_0 + \omega_{2H} \left(S_0^2 + \sum_{n=1}^{\infty} \chi_n^{*}\chi_n \right)\right) \, ,
\end{equation}
where $\omega_H = \sqrt{2 \pi R M_5} \, \lambda_H$ and $\omega_{2H} = \lambda_{2H}$.
The term $H^{\dagger}HS_0$ produces a mixing between $S_0$ and $h$. The states in the gauge eigenstate basis can be written in term of the physical states $H_d$ and $H_s$:
\begin{eqnarray}
h   &=&  H_{d}\, \cos\alpha + H_{s}\, \sin\alpha \nonumber \\
S_0 &=& -H_{d}\, \sin\alpha + H_{s}\, \cos\alpha \, ,
\end{eqnarray}
where $\alpha$ is the mixing angle.
Moreover, the mass term for $h$ and $S_0$, and the coupling $\omega_H$ will depend on the values of $\alpha$ and the masses of the physical states:
\begin{eqnarray}
\label{eq:HSlag}
\omega_H &=& \sin{\left(2 \alpha\right)} \, \frac{m_{H_d}^2 - m_{H_s}^2}{2 v}      \nonumber \\
m_D^2    &=& \cos^2\hspace{-0.2em}\alpha \,  m_{H_d}^2 + \sin^2\hspace{-0.2em}\alpha \, m_{H_s}^2 \nonumber \\
m_S^2    &=& \sin^2\hspace{-0.2em}\alpha \, m_{H_d}^2 + \cos^2\hspace{-0.2em}\alpha \, m_{H_s}^2 \, .
\end{eqnarray}
We fix $m_{H_d} = 125.9$~GeV and $|\sin{(\alpha)}| < 0.2 $ in order to fulfill  Higgs observables.

\section{Coannihilations and relic density}
\label{relic}
In this section, we briefly review the calculation of the relic density of a generic dark matter particle $\chi$ including coannihilation processes from KK particles.
Such a situation occurs when there are particles nearly degenerate in the mass spectrum with $\chi$ (For example~\cite{2006JCAP...12..019P}).
In this case the relic abundance of $\chi$ is determined by its self-annihilation cross section and also by annihilation processes involving heavier particles.
We consider that $\chi$ was in thermal equilibrium in the early universe and also decoupled when it was non relativistic. To estimate the dark matter relic density we have to solve a set of Boltzmann equations which governs the evolution of the number density, $n_{\chi}$:

\begin{eqnarray}
\frac{dn_{\chi}}{dt} + 3Hn_{\chi}=-\langle \sigma_{\mbox{\tiny{eff}}} v\rangle\left(n^2_{\chi}-n^2_{eq}\right),
\end{eqnarray} 
where $H$ is the Hubble expansion rate, $n_{eq}$ is the number density at thermal equilibrium and $\langle \sigma_{\mbox{\tiny{eff}}} v\rangle$ is the thermally averaged effective cross section for annihilation times the relative velocity. The effective annihilation cross section is given by:

\begin{eqnarray}
\label{eq:sigmaeff}
\sigma_{\mbox{\tiny{eff}}}=\sum_{ij=1}^N\sigma_{ij}\frac{g_ig_j}
{g^2_{\mbox{\tiny{eff}}}}(1+\Delta_i)^{3/2}
(1+\Delta_j)^{3/2} e^{-x(\Delta_i+\Delta_j)},
\end{eqnarray}
where $g_i$ is the number of internal degrees of freedom of $\chi_i$, $\sigma_{ij}$ is the cross section for the reaction $\chi_i\chi_j \rightarrow \mbox{SM}$, $\Delta_i= (m_i-m_1)/m_1$ is the mass degeneracy parameter, $m_1$ is the mass of the first KK particle and all  heavier particles have decayed into it, $m_i$ is the mass of the {\it i}-th KK particle with $m_i > m_1$,  $x=m_1/T$, $T$ is the temperature and

\begin{eqnarray}
\label{eq:geff}
g_{\rm eff}=\sum_{i=1}^N g_i(1+\Delta_i)^{3/2}e^{-x\Delta_i} ,
\end{eqnarray}
is the effective number of degrees of freedom including high-order KK particles.

In the case of $s$-wave annihilation, i.e. $\langle \sigma_{\rm eff} v\rangle \propto \sigma_{\rm c} = \, {\rm const.}$, the relic abudance can be estimated by:
\begin{eqnarray}
\label{reldensity}
\Omega_{\chi}h^2 = \frac{1.04\times 10^9 \, {\rm GeV}^{-1} x_F}{g_{\star}^{1/2} M_{pl} \, \displaystyle x_F \int_{x_F}^\infty \sigma_{\rm c} \, x^{-2} dx},
\end{eqnarray}
where $x_F$ is the freeze-out temperature (See \cite{Servant:2002aq} for details about the calculation of $x_F$), $g_{\star}$ counts the number of relativistic degrees of freedom at the freeze-out and $M_{pl}$ is the Planck mass.
In the case of our model, cross sections and internal degrees of freedom are similar for every KK particle.
Assuming that
\begin{eqnarray}
\sigma_{ij} \simeq \sigma_0 \delta_{ij} \, \, {\rm and} \, \,  g_i \simeq g \, ,
\end{eqnarray}
the effective cross section is then:
\begin{eqnarray}
\label{eq:seff}
\sigma_{\rm c} = \sigma_{0} \frac{\displaystyle \sum_{i=1}^n  (1+\Delta_i)^{3} e^{-2 x \Delta_i}}{\displaystyle \left[\sum_{i=1}^n (1+\Delta_i)^{3/2}e^{-x\Delta_i} \right]^2} = \sigma_0 f(x,\Delta_2,n) \, ,
\end{eqnarray}
where the mass splitting between the DM and the {\it i}-th KK particles can be written  in terms of $\Delta_2$:
\begin{equation}
\Delta_i = \sqrt{\displaystyle 1 + \frac{(i^2 - 1)}{3} \Delta_2 (\Delta_2 + 2)} - 1 \, ,
\end{equation}
where $\Delta_2$ ranges between 0 and 1.\\

\begin{figure}[tbp]
	\centering
	\includegraphics[width=\figwidth]{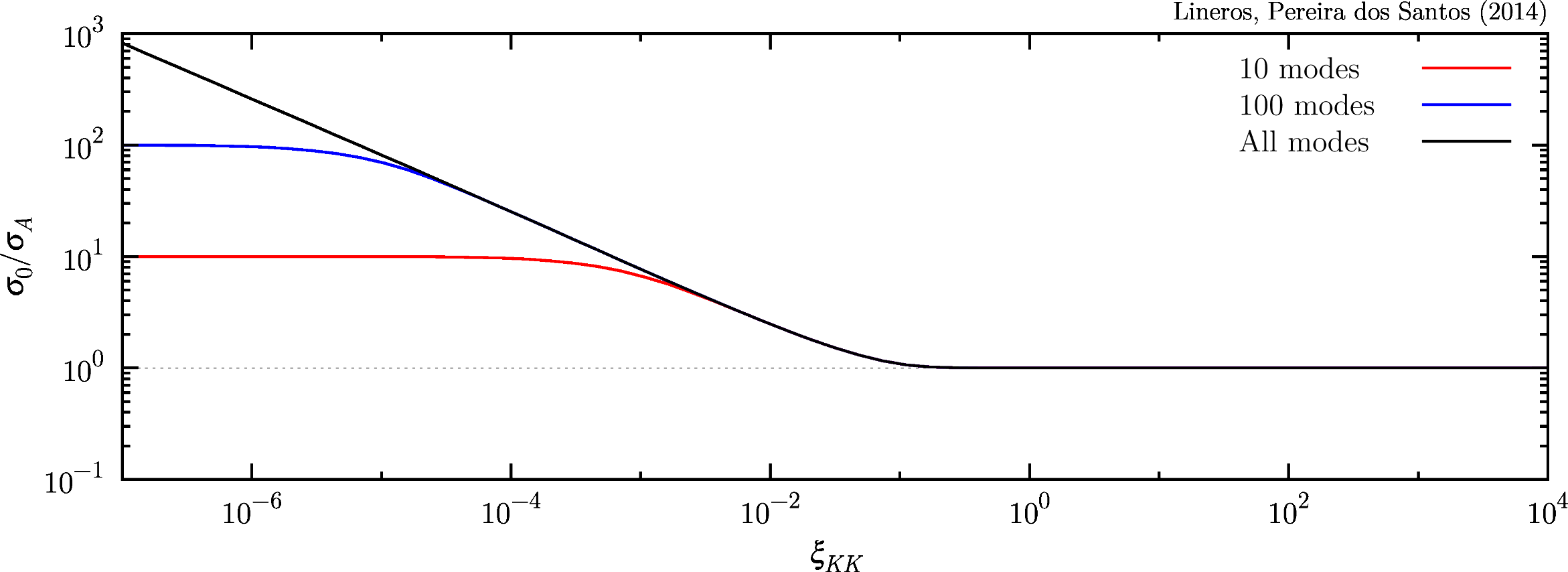} 
	\caption{\label{plots:sled_kkmodes}
Enhancement factor $\sigma_0/\sigma_A$ versus $\xi_{KK}$. The black curve is the convergent limit of \ref{eq:seff}, the blue curve considers the sum up to 100 KK modes, and the red curve up to 10 KK modes.}
\end{figure}

From \ref{reldensity} we can see that the relic abundance depends on a weighted average of the annihilation and coannihilation cross 
sections of all relevant particles and also depends on a mass degeneracy parameter $\Delta_2$.
In order to reproduce the correct value of relic abundance $\sigma_0$ should be:
\begin{equation}
\sigma_0 = \sigma_A \left( x_F \int_{x_F}^{\infty} x^{-2} f(x,\Delta_2,n) dx \right)^{-1} \, ,
\end{equation}
where $\sigma_A$ is the annihilation cross section in the limit with no coannihilations (i.e. the canonical thermal WIMP cross section).
This effect indirectly produces an enhancement into the annihilation cross section.
In figure~\ref{plots:sled_kkmodes}, we present the ratio $\sigma_0/\sigma_A$ in terms of $\xi_{KK}$, where
\begin{equation}
\label{eq:xi}
\xi_{KK} = \frac{2}{\pi} \tan{\left( \frac{\pi}{2} \Delta_{2}\right)} \, ,
\end{equation}
 is a better way to describe the mass splitting in our model.
We note that the enhancement factor depends drastically on the amount of KK modes. Moreover, we show the cases where we truncate the sum in \ref{eq:seff}  including only up to 10 and 100 modes, and the limit $n$-large of $f(x,\Delta_2,n)$.
For small values of $\xi_{KK}$ the truncation induces a fixed amount of enhancement.
In the case $n$-large, the enhancement factor grows continuously when $\xi_{KK} \rightarrow 0$.

We would like to emphasize that we consider $n=10$ KK modes. The reason why we consider this number cames from the current upper bounds from indirect detection searches that are consistent with $\langle \sigma v \rangle \lesssim 10^{-25} \, {\rm cm}^3/{\rm s}$. If we translate these bounds in terms of number of KK modes we find that $n \sim 10$ modes (or mass splitting of $\xi_{KK} > 10^{-3}$).

\section{Results}
\label{results}

\begin{table}[bt]
\centering
\begin{tabular}{|c||c|}
\hline \phantom{xxxxxxxxxx} Parameter \phantom{xxxxxxxxxx} & \phantom{xxxxxxxxxx} Range \phantom{xxxxxxxxxx} \\
\hline \hline
$M_{5}$ (GeV) & 100 --- $10^7$ \\
$M_{KK}$ (GeV) & 1 --- $10^7$ \\

$m_{H_d}$ (GeV) & 125.9 \\
$m_{H_s}$ (GeV) &  1 --- $10^4$\\
$|\sin\alpha|$ & $10^{-2}$  --- 0.2 \\

$\omega_4$ & $10^{-4}$ --- 3 \\
$\omega_3$ & $10^{-4}$ --- $\sqrt{4 \, \omega_4 m_{S}^2}$\\
$\omega_{2H}$ & $10^{-4}$ --- 3 \\
\hline
\end{tabular}
\caption{\label{tab:scan}  Parameter scan ranges. The remaining parameters are calculated from this set.}
\end{table}

In this section we discuss the main results. 
We implement the model using the public codes: {\tt LanHEP}~\cite{2009CoPhC.180..431S} and {\tt MicrOmegas}~\cite{2014CoPhC.185..960B}. As we explained in the last section, here we consider $n=10$ KK modes to be consistent with bounds from indirect detection\footnote{However, the number of scattering amplitudes needed for the relic abundance calculation grows like $n^2$ that imposes technical limitations.}.
%\footnote{If we include more KK modes we will deal with technical (computational) limitations and is possible that the result could be in conflict with current bounds on annihilation cross section.}
%
We perform a scan of the parameter space searching for regions with the correct DM relic abundance~\cite{2013arXiv1303.5076P}:
\begin{equation}
\Omega_{\rm DM} h^2 = 0.1196 \pm 0.0031 \,
\end{equation}
and for SM higgs consistent phenomenology~\cite{2014arXiv1405.3455C}:
\begin{equation}
\label{eq:totwidth}
\Gamma_{h_{\rm SM}} < 0.022 \, {\rm GeV} \, ,
\end{equation} 
as a basic set of constraints. We can safely use \ref{eq:totwidth} in our analysis because $\sin\alpha$ is rather small. This implies that $H_d$ is mainly composed of stardard model Higgs $H$~\cite{2012PhLB..716..179L} (see also \cite{Queiroz:2014yna}).
In addition, we scan on the effective couplings $\omega_i$ avoiding regions with non-perturbative processes.
The numerical range for each parameter are shown in Table~\ref{tab:scan} where the rest of the parameters of the model are calculated using these.
The mass of the lightest KK particle can be written in terms of some input parameters resulting:
\begin{equation}
	m_{S_1} =  \sqrt{\sin^2\hspace{-0.2em}\alpha \, m_{H_d}^2 + \cos^2 \hspace{-0.2em}\alpha \, m_{H_s}^2 + M_{KK}^2} \, ,
\end{equation}
where the scanned DM mass is in the range between $\sqrt{2}$ and $10^7$~GeV. Also it is important to remark that the DM mass receives contributions from the mass of the doublet ($m_{H_d} = 125.9 \, {\rm GeV}$), the mass of the singlet $S_0$ and the contribution from the compactification $M_{KK}$. This implies that DM masses smaller than $\sim 20$~GeV require a very small mixing angle $\sin{\alpha}$ as well as small values of $m_{H_s}$ and $M_{KK}$. The minimum mass for $\sin{\alpha} = 0.2$ would be 25~GeV.
% roberto
This range of masses also has an impact of the Higgs invisible decay, because it is required a small value of $\sin\alpha$ in order to be consistent with the current bounds~\cite{2014arXiv1405.3530B}. On the other hand, leading processes for relic abundance calculation are the ones mediated by $H_s$.\\

When $M_{KK}$ is small with respect to $m_{S}$ (Eq.~\ref{eq:HSlag}), coannihilations become important due to the mass degeneracy among KK states that is less than few percent~\cite{1991PhRvD..43.3191G}.
We use $\xi_{KK}$ (Equation~\ref{eq:xi}) in order to quantify the mass degeneracy and to avoid accumulation of points in the case of $m_{S} \ll M_{KK}$. 
For a fixed value of the DM mass, low values of $\xi_{KK}$ mean that the DM mass has a larger contribution from $m_S$ than from $M_{KK}$.
On the contrary, larger values correspond to larger values of $M_{KK}$.
The parameter $\xi_{KK}$ indicates regions where DM phenomenology is affected by coannihilations.\\
\begin{figure}[tbp]
	\centering
	\includegraphics[width=\figwidth]{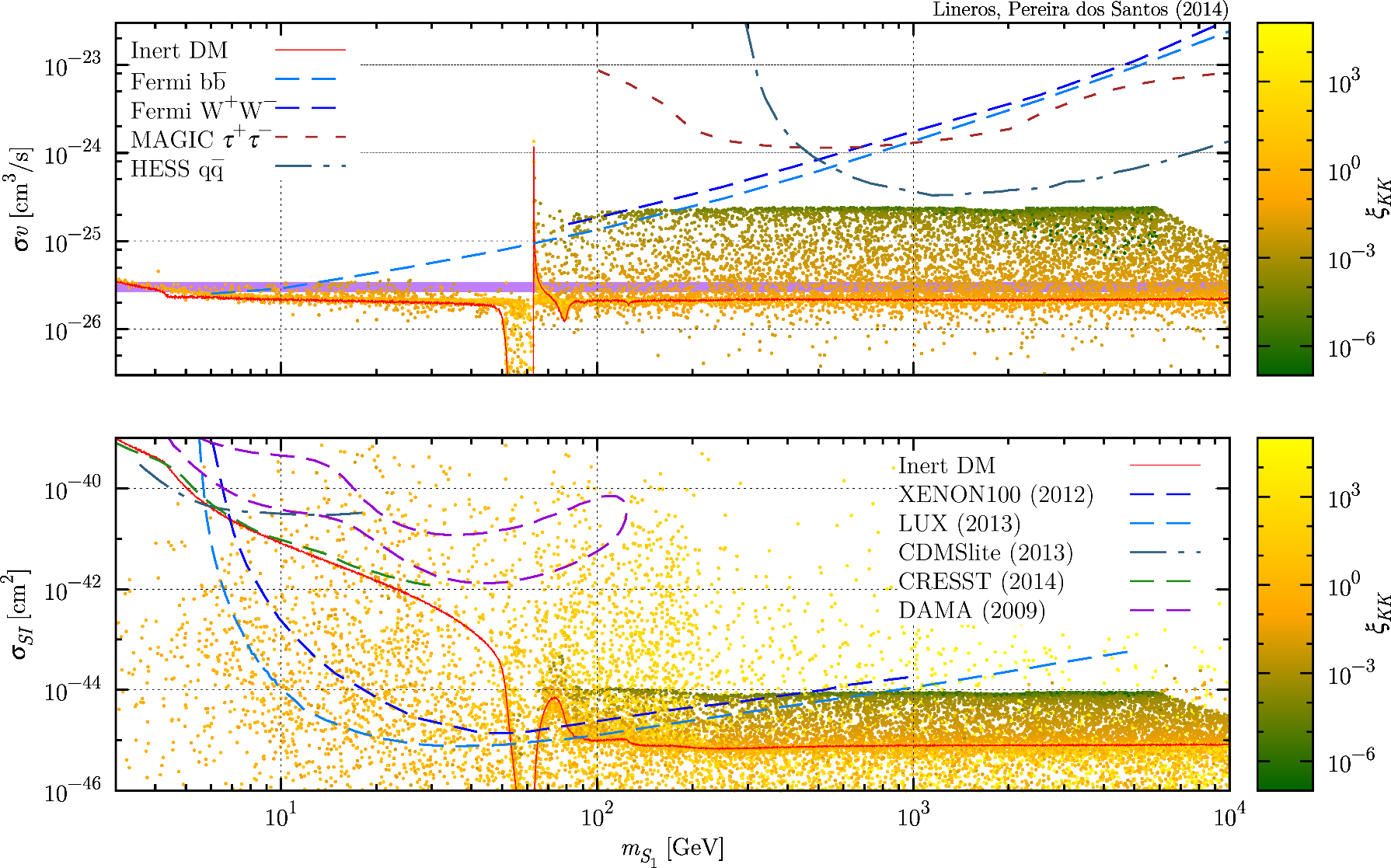} 
	\caption{\label{plots:sled_full}
Top-panel: The annihilation cross section as a function of dark matter mass. The solid red line correponds to the inert singlet DM model's predictions. Bounds from FERMI-LAT~\cite{2013arXiv1310.0828T}, MAGIC~\cite{2013arXiv1312.1535A}, and HESS~\cite{2011PhRvL.106p1301A} are displayed in dashed lines. The color scale correponds to the value of the degeneracy parameter $\xi_{KK}$. Smaller values of $\xi_{KK}$ lead to enhanced annihilations due to more Kaluza Klein modes were involved in the early universe.
Bottom-panel: The spin independent cross section as a function of dark matter mass. Bounds from  XENON100~\cite{XENON:2012Ph}, LUX~\cite{2013arXiv1310.8214L}, CDMSlite~\cite{2013arXiv1309.3259A}, CRESST~\cite{2014arXiv1407.3146C}, and DAMA/LIBRA~\cite{2009JCAP...04..010S} are also shown.}
\end{figure}

In Figure~\ref{plots:sled_full}, we show the annihilation cross section and the spin independent cross section versus the DM mass $m_{S_1}$ for a scan of the models' parameter space. 
The color bar indicates the value of $\xi_{KK}$ where for lower values there is an enhancement of both cross sections due to coannihilations (as in~\cite{1991PhRvD..43.3191G, 2013JCAP...04..044K}).
On the other hand, when $\xi_{KK}$ is large most of the points fall close to the predictions of the inert singlet DM model~\cite{2007JCAP...02..028L}.
Some spread along the inert singlet DM model line is due to that in the limit of large $\xi_{KK}$ the model is the inert DM model with an extra singlet that mixes with the SM higgs.\\

Current bounds from gamma-rays observatories are shown for comparison. We notice that the enhancement due to coannihilations produces points that are just below some of the bounds. This effect makes the model testable in the near future when updated bounds will be released.\\

In the case of direct detection, the effect of coannihiliations also enhances the spin independent cross section for DM masses larger than 70~GeV. 
Here, the enhancement comes from the coupling of $H_d$ with the DM which is correlated with the annihilation cross section.
This correlation is a common feature of models with higgs portal as leading process.
On the other hand, the spin independent cross section is also enhanced when the $S_0$ is very light. This is of special interest for explaining DAMA/LIBRA results~\cite{2009JCAP...04..010S}(see also \cite{Profumo:2014mpa}). \\

\begin{figure}[tbp]
	\centering
	\includegraphics[width=\figwidth]{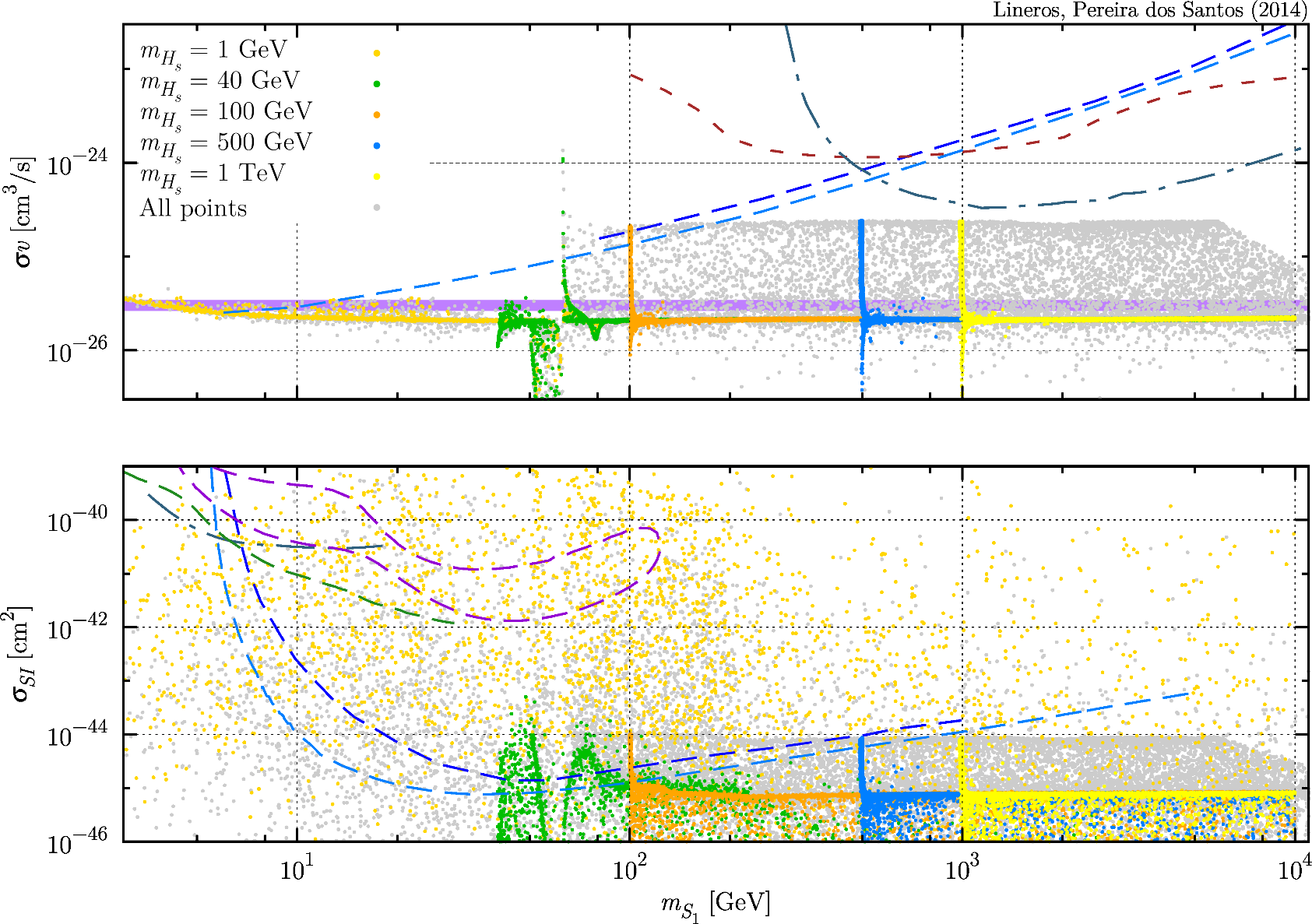} 
	\caption{\label{plots:masses}
Top-panel: Annihilation cross section versus DM mass. Points correspond to scans for fixed values of $m_{H_s}$. An enhancement in the cross section is achieved whenever the DM mass is very close to $m_{H_s}$.
Bottom-panel: Spin independent cross section versus DM mass. The enhancement due to coannihiliations is also present. For $m_{H_s} = 1 \, {\rm GeV}$, the enhancement is due to $m_{H_s}$ lightness and it is present in the whole DM mass range.
Observational bounds are shown and are the same as in Figure~\ref{plots:sled_full}}. 
\end{figure}

We analyze the case in which we scan on the DM mass keeping $m_{H_s}$ fixed. This is shown in Figure~\ref{plots:masses} where we consider 5 fixed values of $m_{H_s}$ to calculate the annihilation cross section and the spin independent cross section.
As was explained before, small values of $M_{KK}$ with respect to $m_{S}$ lead to a degenerate spectrum and therefore coannihilations effect are important and produce an enhanced cross section value. 
The cases, where this effect cannot be appreciated, correspond to $m_{H_s} = 1 \, , \,40 \, {\rm GeV}$ because of the lightness of $m_{H_s}$ and the presence of resonances for DM masses below 90~GeV. \\
\begin{figure}[tbp]
	\centering
	\includegraphics[width=\figwidth]{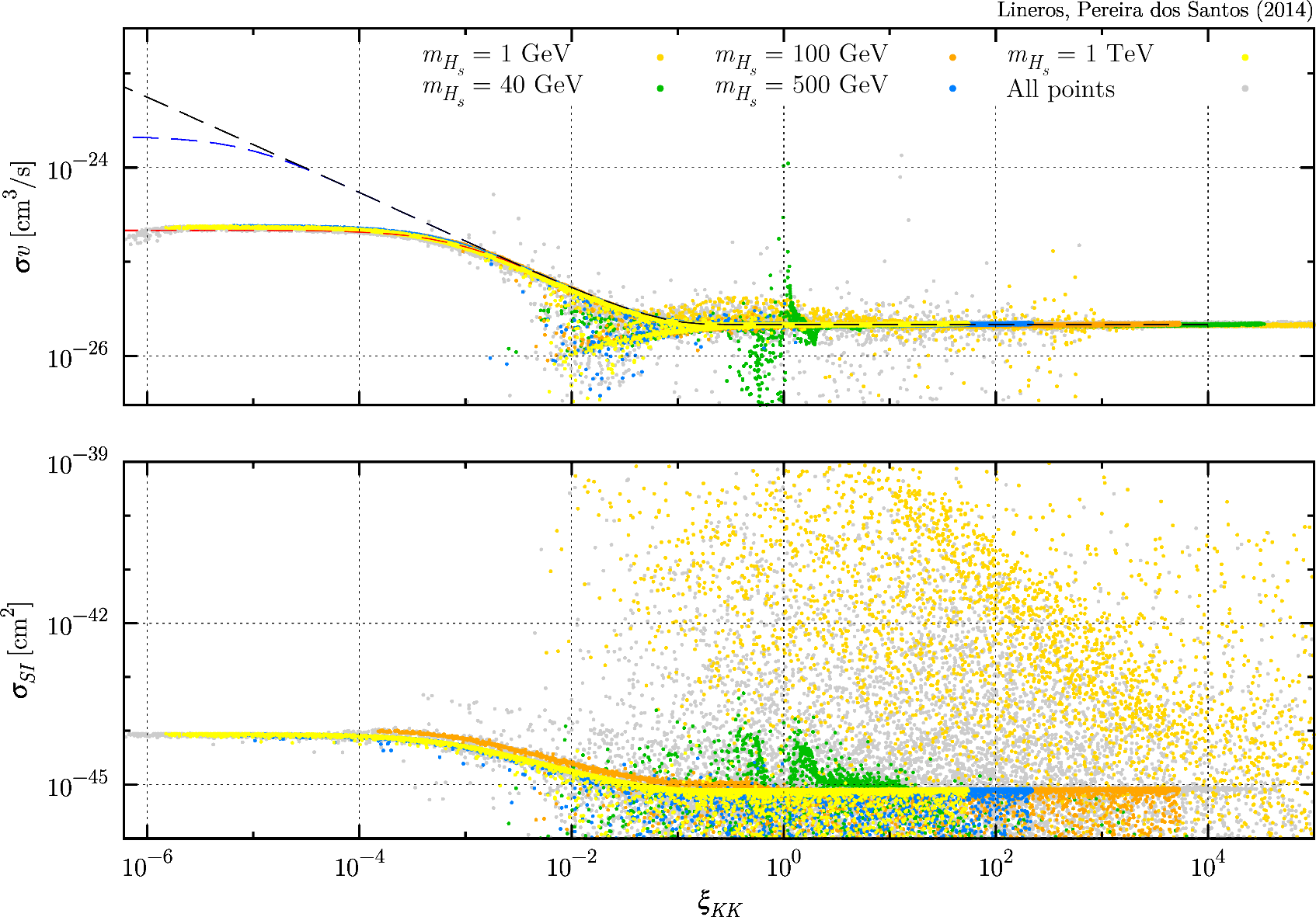} 
	\caption{	\label{plots:xiK}
Annihilation cross section (top) and Spin Independent cross section (bottom) versus $\xi_{KK}$. Both cross sections converge to a fixed value for low values of $\xi_{KK}$. Dashed curves correspond to ones in Figure~\ref{plots:sled_kkmodes}}
\end{figure}

Figure~\ref{plots:xiK} shows the dependence in terms of $\xi_{KK}$ of both the annihilation cross section and the spin independent cross section.
Similar behavior is present in both cross sections for lower values of $\xi_{KK}$. For $\xi_{KK} < 10^{-2}$, i.e. mass splitting of the order of 1\% of the DM mass, each cross section starts to converge to a common value: $\sigma_{\rm SI} \simeq 10^{-44} \, {\rm cm}^2$ and $\sigma v \simeq 2 \times 10^{-25} \, {\rm cm}^3/{\rm s}$. In fact, this is compatible with the effect of coannihilations discussed in the previous section.
We also confront the enhancement factor $\sigma_0/\sigma_A$ with the scan results. We observe that the curve for $n=10$ is fully compatible with our results.\\
\begin{figure}[tbp]
	\centering
	\includegraphics[width=\figwidth]{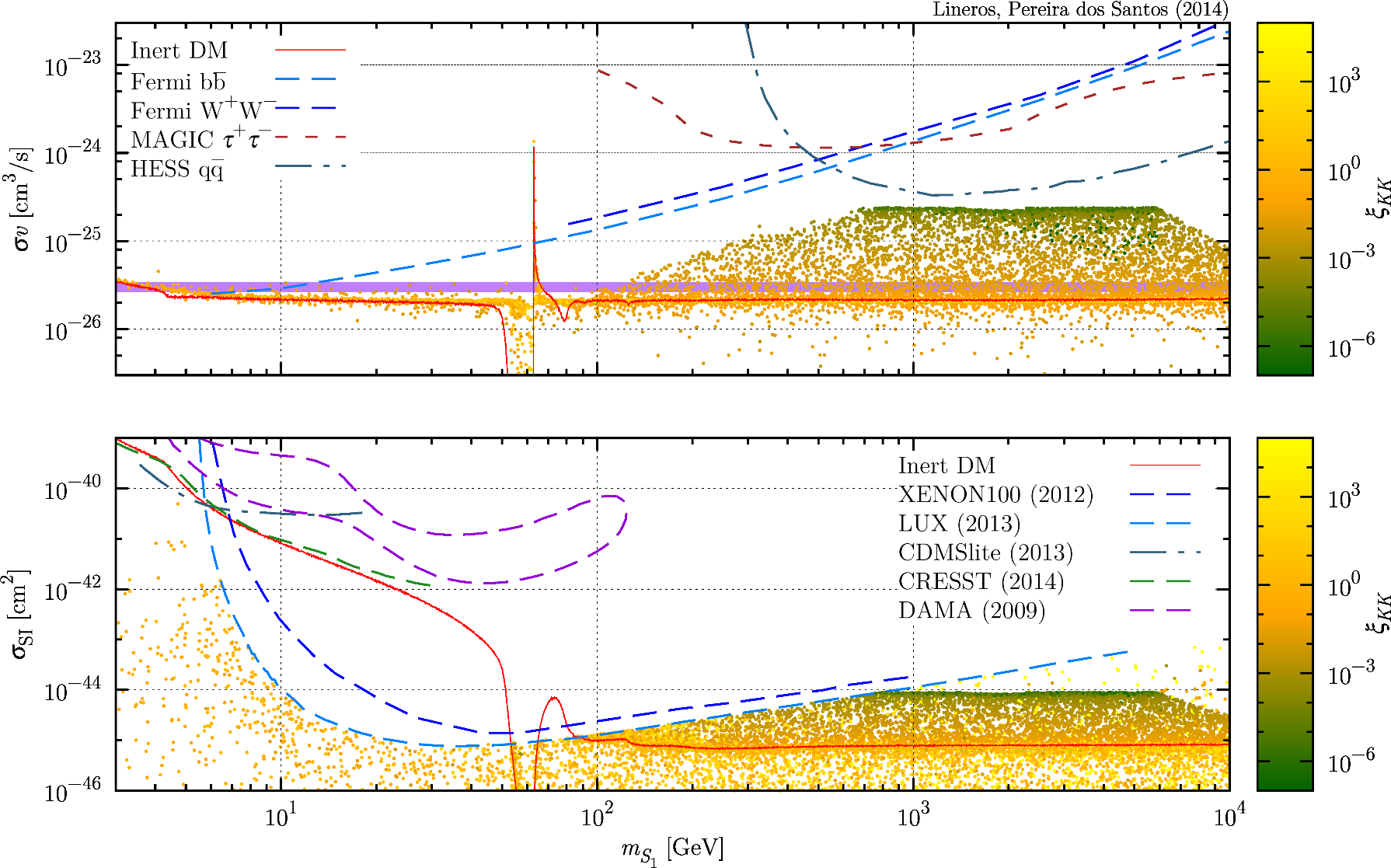} 
	\caption{\label{plots:lux}
Annihilation cross section (top-panel) and Spin Independent cross section (bottom-panel) vs DM mass excluding points that do not satisfy latest LUX bounds~\cite{2013arXiv1310.8214L}. }
\end{figure}

\emph{Including LUX bounds:}
The LUX collaboration~\cite{2013arXiv1310.8214L} has set one of the strongest bounds for the WIMP--nucleon cross section.
In figure~\ref{plots:lux}, we present a scan that includes the constraints from LUX experiment.
As is expected, these bounds have enormous impact on the region allowed by cosmology and higgs width. Also, as we can see from figure \ref{plots:lux} the allowed region for spin independent cross section is significantly reduced.
We highlight that the indirect detection signal still presents an enhancement due to coannihilations after imposing LUX constraints. 
Compared to the inert DM model, our model produces an increased signal for DM masses larger than 100~GeV. This feature allows to test it in current gamma-rays observatories like FERMI-LAT and the Cerenkov Telescope Array (CTA)~\cite{2013APh....43..189D}.
Compared with the inert higgs DM model, our model can be probed in the near future by newer analyses of LUX or XENON1T~\cite{2012arXiv1206.6288A}.\\

\begin{figure}[tbp]
	\centering
	\includegraphics[width=\figwidth]{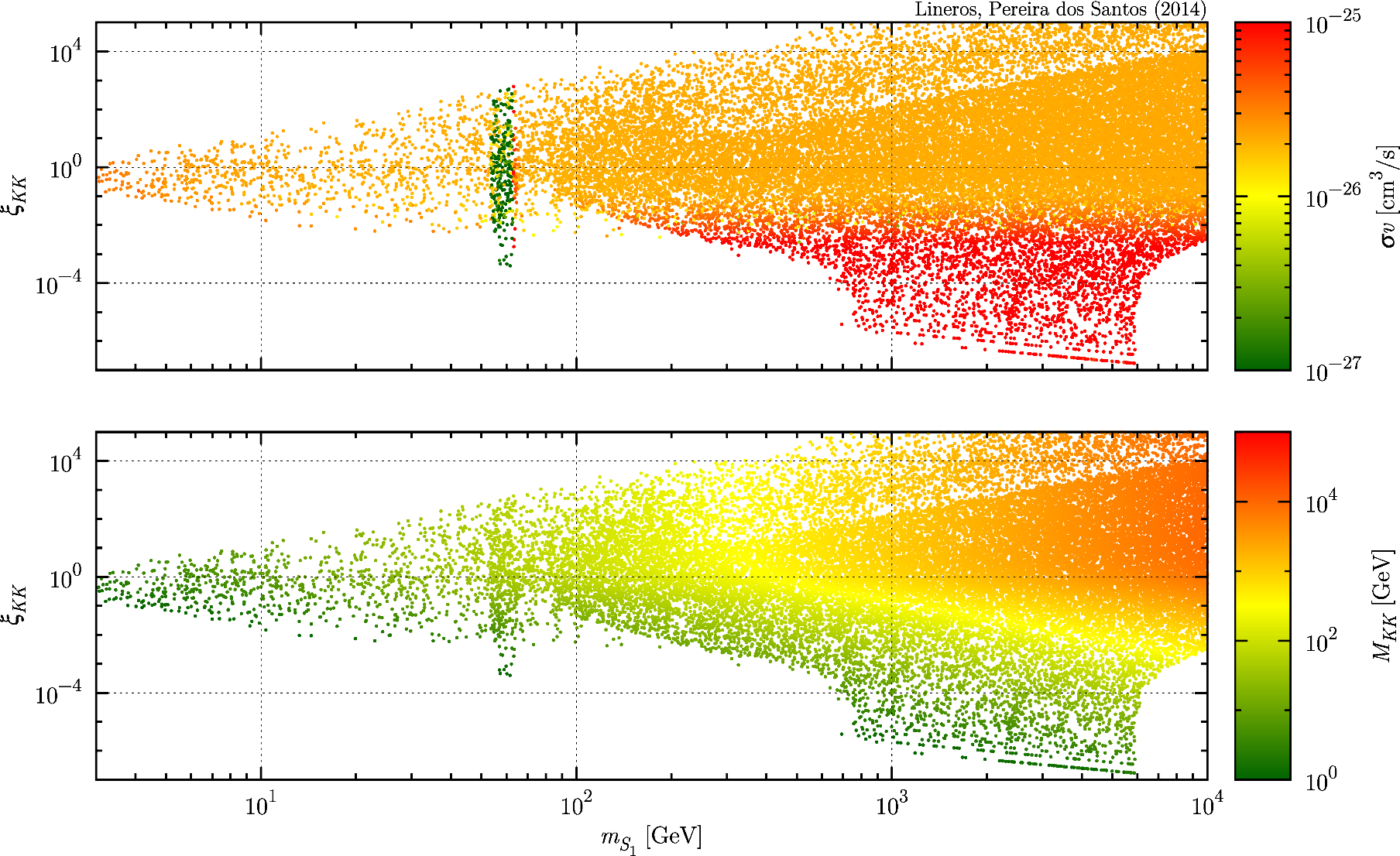}
	\caption{\label{plots:xi_versus}
Top-panel: $\xi_{KK}$ versus DM mass. Color scale corresponds to the annihilation cross section value. 
Bottom-panel: $\xi_{KK}$ versus DM mass. Color scale corresponds to the value of $M_{KK}$
}

\end{figure}

In Figure~\ref{plots:xi_versus}, we show the plane $\xi_{KK}$ vs DM mass including the annihilation cross section and $M_{KK}$ both in color scale.
Here, we observe that points with $ \langle \sigma v \rangle > 3 \times 10^{-26}\, {\rm cm}^3{\rm s}^{-1}$ are in the range: $150 \, {\rm GeV} \lesssim m_{S_1} \lesssim 10 \, {\rm TeV}$. 
These points also satisfies that $M_{KK} \lesssim 10^{2}~{\rm GeV}$.
The LUX constraints appeared as an empty zone in $80 \, {\rm GeV} \lesssim m_{S_1} \lesssim 700 \, {\rm GeV}$.

\section{Conclusions}
\label{sec:concl}

The origin of Dark Matter is still puzzling the scientific community.
In this work, we study a DM model inspired by extra dimension theories.
The DM candidate is the lightest KK particle that emerges from the compactification of a SM singlet real scalar existing in the (4+1) dimensions.
We obtain the effective model at (3+1) dimensions by integrating over the extra dimension, and then we calculate the indirect and direct detection cross section requiring correct value for DM relic abundance and SM higgs phenomenology.\\
The model contains nice features which allow us to study the effect of coannihilations when several particles are quasi degenerate.
We found that for a degenerate mass spectrum both annihilation and spin independent cross section converge to fixed values which are larger than for canonical WIMPs.
We consider up to 10 KK modes in our analysis.
In the case of the annihilation cross section, it converges to a value of $2 \times 10^{-25} \, {\rm cm}^3 / {\rm s}$.
This enhancement in the cross section is translated to a mass splitting among the KK particles smaller than 100~GeV. 
For larger mass splittings, the model converges to the predictions of the inert scalar DM model.\\

We would like to point out that because of the presence of coannihilations, as we can see from our results,
 for small $\xi_{KK}$ the size of indirect and direct detection signal provides a test for the model in the near future.
In principle, it may be proved by confronting it with new data from FERMI-LAT, LUX or XENON1T.

\section*{Acknowledgments}

We thank Nicolao Fornengo, Martin Hirsch, and Carlos Yaguna for discussions and comments.
This work was supported by the Spanish MINECO under grants FPA2011-22975 and MULTIDARK CSD2009-00064 (Consolider-Ingenio 2010 Programme), by Prometeo/2009/091 (Generalitat Valenciana) and by Brazilian research council CNPq.
 R.L. is supported by a Juan de la Cierva contract (MINECO). 
%\nocite{*}
%%%%%%%% Journals %%%%%%%%%%%%%%
\def\apjl{Astrophys.~J.~Lett.}                % Astrophysical Journal, Letters
\def\apjs{Astrophys.~J.~Suppl.~Ser.}          % Astrophysical Journal, Supplement
\def\aap{Astron.~\&~Astrophys.}               % Astronomy and Astrophysics
\def\aj{Astron.~J.}                           %
\def\araa{Ann.~Rev.~Astron.~Astrophys.}       %
\def\mnras{Mon.~Not.~R.~Astron.~Soc.}         %
\def\physrep{Phys.~Rept.}                     %
\def\jcap{J.~Cosmology~Astropart.~Phys.}      % Journal of Cosmology and Astroparticle Physics
\def\jhep{J.~High~Ener.~Phys.}                % Journal of High Energy Physics
\def\prl{Phys.~Rev.~Lett.}                    % Physical Review Letters
\def\prd{Phys.~Rev.~D}                        % Physical Review D
\def\nphysa{Nucl.~Phys.~A}                    % Nuclear Physics A

\bibliographystyle{JHEP}
\bibliography{refpaper.bib}

\providecommand{\href}[2]{#2}\begingroup\raggedright\begin{thebibliography}{10}

\bibitem{Bertone:2004pz}
G.~Bertone, D.~Hooper, and J.~Silk, {\it {Particle dark matter: Evidence,
  candidates and constraints}},  {\em Phys.Rept.} {\bf 405} (2005) 279--390,
  [\href{http://arxiv.org/abs/hep-ph/0404175}{{\tt hep-ph/0404175}}].

\bibitem{Jungman:1995df}
G.~Jungman, M.~Kamionkowski, and K.~Griest, {\it {Supersymmetric dark matter}},
   {\em Phys.Rept.} {\bf 267} (1996) 195--373,
  [\href{http://arxiv.org/abs/hep-ph/9506380}{{\tt hep-ph/9506380}}].

\bibitem{Peccei:1977hh}
R.~Peccei and H.~R. Quinn, {\it {CP Conservation in the Presence of
  Instantons}},  {\em Phys.Rev.Lett.} {\bf 38} (1977) 1440--1443.

\bibitem{Peccei:1977ur}
R.~Peccei and H.~R. Quinn, {\it {Constraints Imposed by CP Conservation in the
  Presence of Instantons}},  {\em Phys.Rev.} {\bf D16} (1977) 1791--1797.

\bibitem{Weinberg:1977ma}
S.~Weinberg, {\it {A New Light Boson?}},  {\em Phys.Rev.Lett.} {\bf 40} (1978)
  223--226.

\bibitem{Wilczek:1977pj}
F.~Wilczek, {\it {Problem of Strong p and t Invariance in the Presence of
  Instantons}},  {\em Phys.Rev.Lett.} {\bf 40} (1978) 279--282.

\bibitem{Dine:1982ah}
M.~Dine and W.~Fischler, {\it {The Not So Harmless Axion}},  {\em Phys.Lett.}
  {\bf B120} (1983) 137--141.

\bibitem{Abbott:1982af}
L.~Abbott and P.~Sikivie, {\it {A Cosmological Bound on the Invisible Axion}},
  {\em Phys.Lett.} {\bf B120} (1983) 133--136.

\bibitem{Preskill:1982cy}
J.~Preskill, M.~B. Wise, and F.~Wilczek, {\it {Cosmology of the Invisible
  Axion}},  {\em Phys.Lett.} {\bf B120} (1983) 127--132.

\bibitem{ArkaniHamed:1998rs}
N.~Arkani-Hamed, S.~Dimopoulos, and G.~Dvali, {\it {The Hierarchy problem and
  new dimensions at a millimeter}},  {\em Phys.Lett.} {\bf B429} (1998)
  263--272, [\href{http://arxiv.org/abs/hep-ph/9803315}{{\tt hep-ph/9803315}}].

\bibitem{Antoniadis:1998ig}
I.~Antoniadis, N.~Arkani-Hamed, S.~Dimopoulos, and G.~Dvali, {\it {New
  dimensions at a millimeter to a Fermi and superstrings at a TeV}},  {\em
  Phys.Lett.} {\bf B436} (1998) 257--263,
  [\href{http://arxiv.org/abs/hep-ph/9804398}{{\tt hep-ph/9804398}}].

\bibitem{Randall:1999vf}
L.~Randall and R.~Sundrum, {\it {An Alternative to compactification}},  {\em
  Phys.Rev.Lett.} {\bf 83} (1999) 4690--4693,
  [\href{http://arxiv.org/abs/hep-th/9906064}{{\tt hep-th/9906064}}].

\bibitem{Randall:1999ee}
L.~Randall and R.~Sundrum, {\it {A Large mass hierarchy from a small extra
  dimension}},  {\em Phys.Rev.Lett.} {\bf 83} (1999) 3370--3373,
  [\href{http://arxiv.org/abs/hep-ph/9905221}{{\tt hep-ph/9905221}}].

\bibitem{Servant:2002aq}
G.~Servant and T.~M. Tait, {\it {Is the lightest Kaluza-Klein particle a viable
  dark matter candidate?}},  {\em Nucl.Phys.} {\bf B650} (2003) 391--419,
  [\href{http://arxiv.org/abs/hep-ph/0206071}{{\tt hep-ph/0206071}}].

\bibitem{Cheng:2002ej}
H.-C. Cheng, J.~L. Feng, and K.~T. Matchev, {\it {Kaluza-Klein dark matter}},
  {\em Phys.Rev.Lett.} {\bf 89} (2002) 211301,
  [\href{http://arxiv.org/abs/hep-ph/0207125}{{\tt hep-ph/0207125}}].

\bibitem{Ponton:2008zv}
E.~Ponton and L.~Randall, {\it {TeV Scale Singlet Dark Matter}},  {\em JHEP}
  {\bf 0904} (2009) 080, [\href{http://arxiv.org/abs/0811.1029}{{\tt
  arXiv:0811.1029}}].

\bibitem{Medina:2011qc}
A.~D. Medina and E.~Ponton, {\it {Warped Radion Dark Matter}},  {\em JHEP} {\bf
  1109} (2011) 016, [\href{http://arxiv.org/abs/1104.4124}{{\tt
  arXiv:1104.4124}}].

\bibitem{Appelquist:2000nn}
T.~Appelquist, H.-C. Cheng, and B.~A. Dobrescu, {\it {Bounds on universal extra
  dimensions}},  {\em Phys.Rev.} {\bf D64} (2001) 035002,
  [\href{http://arxiv.org/abs/hep-ph/0012100}{{\tt hep-ph/0012100}}].

\bibitem{Georgi:2000ks}
H.~Georgi, A.~K. Grant, and G.~Hailu, {\it {Brane couplings from bulk loops}},
  {\em Phys.Lett.} {\bf B506} (2001) 207--214,
  [\href{http://arxiv.org/abs/hep-ph/0012379}{{\tt hep-ph/0012379}}].

\bibitem{Burnell:2005hm}
F.~Burnell and G.~D. Kribs, {\it {The Abundance of Kaluza-Klein dark matter
  with coannihilation}},  {\em Phys.Rev.} {\bf D73} (2006) 015001,
  [\href{http://arxiv.org/abs/hep-ph/0509118}{{\tt hep-ph/0509118}}].

\bibitem{Deshpande:1977rw}
N.~G. Deshpande and E.~Ma, {\it {Pattern of Symmetry Breaking with Two Higgs
  Doublets}},  {\em Phys.Rev.} {\bf D18} (1978) 2574.

\bibitem{Silveira:1985rk}
V.~Silveira and A.~Zee, {\it {SCALAR PHANTOMS}},  {\em Phys.Lett.} {\bf B161}
  (1985) 136.

\bibitem{McDonald:1993ex}
J.~McDonald, {\it {Gauge singlet scalars as cold dark matter}},  {\em
  Phys.Rev.} {\bf D50} (1994) 3637--3649,
  [\href{http://arxiv.org/abs/hep-ph/0702143}{{\tt hep-ph/0702143}}].

\bibitem{Barger:2008jx}
V.~Barger, P.~Langacker, M.~McCaskey, M.~Ramsey-Musolf, and G.~Shaughnessy,
  {\it {Complex Singlet Extension of the Standard Model}},  {\em Phys.Rev.}
  {\bf D79} (2009) 015018, [\href{http://arxiv.org/abs/0811.0393}{{\tt
  arXiv:0811.0393}}].

\bibitem{Burgess:2000yq}
C.~Burgess, M.~Pospelov, and T.~ter Veldhuis, {\it {The Minimal model of
  nonbaryonic dark matter: A Singlet scalar}},  {\em Nucl.Phys.} {\bf B619}
  (2001) 709--728, [\href{http://arxiv.org/abs/hep-ph/0011335}{{\tt
  hep-ph/0011335}}].

\bibitem{2013arXiv1310.0828T}
{The Fermi-LAT Collaboration}, {\it {Dark Matter Constraints from Observations
  of 25 Milky Way Satellite Galaxies with the Fermi Large Area Telescope}},
  {\em ArXiv e-prints} (Oct., 2013) [\href{http://arxiv.org/abs/1310.0828}{{\tt
  arXiv:1310.0828}}].

\bibitem{2013arXiv1312.1535A}
J.~{Aleksi{\'c}} et~al., {\it {Optimized dark matter searches in deep
  observations of Segue 1 with MAGIC}},  {\em ArXiv e-prints} (Dec., 2013)
  [\href{http://arxiv.org/abs/1312.1535}{{\tt arXiv:1312.1535}}].

\bibitem{2011PhRvL.106p1301A}
A.~{Abramowski} et~al., {\it {Search for a Dark Matter Annihilation Signal from
  the Galactic Center Halo with H.E.S.S.}},  {\em Physical Review Letters} {\bf
  106} (Apr., 2011) 161301, [\href{http://arxiv.org/abs/1103.3266}{{\tt
  arXiv:1103.3266}}].

\bibitem{XENON:2012Ph}
E.~{Aprile} et~al., {\it {Dark Matter Results from 225 Live Days of XENON100
  Data}},  {\em Physical Review Letters} {\bf 109} (Nov., 2012) 181301,
  [\href{http://arxiv.org/abs/1207.5988}{{\tt arXiv:1207.5988}}].

\bibitem{2013arXiv1310.8214L}
{LUX Collaboration}, {\it {First results from the LUX dark matter experiment at
  the Sanford Underground Research Facility}},  {\em ArXiv e-prints} (Oct.,
  2013) [\href{http://arxiv.org/abs/1310.8214}{{\tt arXiv:1310.8214}}].

\bibitem{2013arXiv1309.3259A}
R.~{Agnese} et~al., {\it {CDMSlite: A Search for Low-Mass WIMPs using
  Voltage-Assisted Calorimetric Ionization Detection in the SuperCDMS
  Experiment}},  {\em ArXiv e-prints} (Sept., 2013)
  [\href{http://arxiv.org/abs/1309.3259}{{\tt arXiv:1309.3259}}].

\bibitem{2014arXiv1407.3146C}
{CRESST Collaboration}, G.~{Angloher}, A.~{Bento}, C.~{Bucci}, L.~{Canonica},
  A.~{Erb}, F.~v. {Feilitzsch}, N.~{Ferreiro Iachellini}, P.~{Gorla},
  A.~{G{\"u}tlein}, D.~{Hauff}, P.~{Huff}, J.~{Jochum}, M.~{Kiefer},
  C.~{Kister}, H.~{Kluck}, H.~{Kraus}, J.-C. {Lanfranchi}, J.~{Loebell},
  A.~{M{\"u}nster}, F.~{Petricca}, W.~{Potzel}, F.~{Pr{\"o}bst}, F.~{Reindl},
  S.~{Roth}, K.~{Rottler}, C.~{Sailer}, K.~{Sch{\"a}ffner}, J.~{Schieck},
  J.~{Schmaler}, S.~{Scholl}, S.~{Sch{\"o}nert}, W.~{Seidel}, M.~v. {Sivers},
  L.~{Stodolsky}, C.~{Strandhagen}, R.~{Strauss}, A.~{Tanzke}, M.~{Uffinger},
  A.~{Ulrich}, I.~{Usherov}, M.~{W{\"u}strich}, S.~{Wawoczny}, M.~{Willers},
  and A.~{Z{\"o}ller}, {\it {Results on low mass WIMPs using an upgraded
  CRESST-II detector}},  {\em ArXiv e-prints} (July, 2014)
  [\href{http://arxiv.org/abs/1407.3146}{{\tt arXiv:1407.3146}}].

\bibitem{2009JCAP...04..010S}
C.~{Savage}, G.~{Gelmini}, P.~{Gondolo}, and K.~{Freese}, {\it {Compatibility
  of DAMA/LIBRA dark matter detection with other searches}},  {\em \jcap} {\bf
  4} (Apr., 2009) 10, [\href{http://arxiv.org/abs/0808.3607}{{\tt
  arXiv:0808.3607}}].

\bibitem{Cheng:1999bg}
H.-C. Cheng, B.~A. Dobrescu, and C.~T. Hill, {\it {Electroweak symmetry
  breaking and extra dimensions}},  {\em Nucl.Phys.} {\bf B589} (2000)
  249--268, [\href{http://arxiv.org/abs/hep-ph/9912343}{{\tt hep-ph/9912343}}].

\bibitem{Biggio:2003kp}
C.~Biggio, F.~Feruglio, I.~Masina, and M.~Perez-Victoria, {\it {Fermion
  generations, masses and mixing angles from extra dimensions}},  {\em
  Nucl.Phys.} {\bf B677} (2004) 451--470,
  [\href{http://arxiv.org/abs/hep-ph/0305129}{{\tt hep-ph/0305129}}].

\bibitem{1999PhRvL..83.4245C}
J.~M. {Cline}, C.~{Grojean}, and G.~{Servant}, {\it {Cosmological Expansion in
  the Presence of an Extra Dimension}},  {\em Physical Review Letters} {\bf 83}
  (Nov., 1999) 4245--4248, [\href{http://arxiv.org/abs/hep-ph/9906523}{{\tt
  hep-ph/9906523}}].

\bibitem{Ponton:2012bi}
E.~Ponton, {\it {TASI 2011: Four Lectures on TeV Scale Extra Dimensions}},
  \href{http://arxiv.org/abs/1207.3827}{{\tt arXiv:1207.3827}}.

\bibitem{Antoniadis:1990ew}
I.~Antoniadis, {\it {A Possible new dimension at a few TeV}},  {\em Phys.Lett.}
  {\bf B246} (1990) 377--384.

\bibitem{ArkaniHamed:1998nn}
N.~Arkani-Hamed, S.~Dimopoulos, and G.~Dvali, {\it {Phenomenology, astrophysics
  and cosmology of theories with submillimeter dimensions and TeV scale quantum
  gravity}},  {\em Phys.Rev.} {\bf D59} (1999) 086004,
  [\href{http://arxiv.org/abs/hep-ph/9807344}{{\tt hep-ph/9807344}}].

\bibitem{Dienes:1998vg}
K.~R. Dienes, E.~Dudas, and T.~Gherghetta, {\it {Grand unification at
  intermediate mass scales through extra dimensions}},  {\em Nucl.Phys.} {\bf
  B537} (1999) 47--108, [\href{http://arxiv.org/abs/hep-ph/9806292}{{\tt
  hep-ph/9806292}}].

\bibitem{Georgi:2000wb}
H.~Georgi, A.~K. Grant, and G.~Hailu, {\it {Chiral fermions, orbifolds, scalars
  and fat branes}},  {\em Phys.Rev.} {\bf D63} (2001) 064027,
  [\href{http://arxiv.org/abs/hep-ph/0007350}{{\tt hep-ph/0007350}}].

\bibitem{Datta:2012xy}
A.~Datta, U.~K. Dey, A.~Shaw, and A.~Raychaudhuri, {\it {Universal
  Extra-Dimensional Models with Boundary Localized Kinetic Terms: Probing at
  the LHC}},  {\em Phys.Rev.} {\bf D87} (2013) 076002,
  [\href{http://arxiv.org/abs/1205.4334}{{\tt arXiv:1205.4334}}].

\bibitem{Datta:2013nua}
A.~Datta, U.~K. Dey, A.~Raychaudhuri, and A.~Shaw, {\it {Boundary Localized
  Terms in Universal Extra-Dimensional Models through a Dark Matter
  perspective}},  {\em Phys.Rev.} {\bf D88} (2013) 016011,
  [\href{http://arxiv.org/abs/1305.4507}{{\tt arXiv:1305.4507}}].

\bibitem{2006JCAP...12..019P}
S.~{Profumo} and A.~{Provenza}, {\it {Increasing the neutralino relic abundance
  with slepton coannihilations: consequences for indirect dark matter
  detection}},  {\em \jcap} {\bf 12} (Dec., 2006) 19,
  [\href{http://arxiv.org/abs/hep-ph/0609290}{{\tt hep-ph/0609290}}].

\bibitem{2009CoPhC.180..431S}
A.~V. {Semenov}, {\it {LanHEP - a package for the automatic generation of
  Feynman rules in field theory. Version 3.0}},  {\em Computer Physics
  Communications} {\bf 180} (Mar., 2009) 431--454,
  [\href{http://arxiv.org/abs/0805.0555}{{\tt arXiv:0805.0555}}].

\bibitem{2014CoPhC.185..960B}
G.~{B{\'e}langer}, F.~{Boudjema}, A.~{Pukhov}, and A.~{Semenov}, {\it
  {micrOMEGAs 3: A program for calculating dark matter observables}},  {\em
  Computer Physics Communications} {\bf 185} (Mar., 2014) 960--985,
  [\href{http://arxiv.org/abs/1305.0237}{{\tt arXiv:1305.0237}}].

\bibitem{2013arXiv1303.5076P}
{Planck Collaboration}, P.~A.~R. {Ade}, N.~{Aghanim}, C.~{Armitage-Caplan},
  M.~{Arnaud}, M.~{Ashdown}, F.~{Atrio-Barandela}, J.~{Aumont},
  C.~{Baccigalupi}, A.~J. {Banday}, and et~al., {\it {Planck 2013 results. XVI.
  Cosmological parameters}},  {\em ArXiv e-prints} (Mar., 2013)
  [\href{http://arxiv.org/abs/1303.5076}{{\tt arXiv:1303.5076}}].

\bibitem{2014arXiv1405.3455C}
{CMS Collaboration}, {\it {Constraints on the Higgs boson width from off-shell
  production and decay to Z-boson pairs}},  {\em ArXiv e-prints} (May, 2014)
  [\href{http://arxiv.org/abs/1405.3455}{{\tt arXiv:1405.3455}}].

\bibitem{2012PhLB..716..179L}
L.~{Lopez-Honorez}, T.~{Schwetz}, and J.~{Zupan}, {\it {Higgs portal, fermionic
  dark matter, and a Standard Model like Higgs at 125 GeV}},  {\em Physics
  Letters B} {\bf 716} (Sept., 2012) 179--185,
  [\href{http://arxiv.org/abs/1203.2064}{{\tt arXiv:1203.2064}}].

\bibitem{Queiroz:2014yna}
F.~S. Queiroz and K.~Sinha, {\it {The Poker Face of the Majoron Dark Matter
  Model: LUX to keV Line}},  {\em Phys.Lett.} {\bf B735} (2014) 69--74,
  [\href{http://arxiv.org/abs/1404.1400}{{\tt arXiv:1404.1400}}].

\bibitem{2014arXiv1405.3530B}
S.~{Baek}, P.~{Ko}, and W.-I. {Park}, {\it {Invisible Higgs Decay Width vs.
  Dark Matter Direct Detection Cross Section in Higgs Portal Dark Matter
  Models}},  {\em ArXiv e-prints} (May, 2014)
  [\href{http://arxiv.org/abs/1405.3530}{{\tt arXiv:1405.3530}}].

\bibitem{1991PhRvD..43.3191G}
K.~{Griest} and D.~{Seckel}, {\it {Three exceptions in the calculation of relic
  abundances}},  {\em \prd} {\bf 43} (May, 1991) 3191--3203.

\bibitem{2013JCAP...04..044K}
M.~{Klasen}, C.~E. {Yaguna}, J.~D. {Ruiz-{\'A}lvarez}, D.~{Restrepo}, and
  O.~{Zapata}, {\it {Scalar dark matter and fermion coannihilations in the
  radiative seesaw model}},  {\em \jcap} {\bf 4} (Apr., 2013) 44,
  [\href{http://arxiv.org/abs/1302.5298}{{\tt arXiv:1302.5298}}].

\bibitem{2007JCAP...02..028L}
L.~{Lopez Honorez}, E.~{Nezri}, J.~F. {Oliver}, and M.~H.~G. {Tytgat}, {\it
  {The inert doublet model: an archetype for dark matter}},  {\em \jcap} {\bf
  2} (Feb., 2007) 28, [\href{http://arxiv.org/abs/hep-ph/0612275}{{\tt
  hep-ph/0612275}}].

\bibitem{Profumo:2014mpa}
S.~Profumo and F.~S. Queiroz, {\it {GeV WIMPs scattering off of OH impurities
  cannot explain the DAMA signal}},  {\em JCAP} {\bf 1405} (2014) 038,
  [\href{http://arxiv.org/abs/1401.4253}{{\tt arXiv:1401.4253}}].

\bibitem{2013APh....43..189D}
M.~{Doro}, J.~{Conrad}, D.~{Emmanoulopoulos}, M.~A. {S{\`a}nchez-Conde}, J.~A.
  {Barrio}, E.~{Birsin}, J.~{Bolmont}, P.~{Brun}, S.~{Colafrancesco}, S.~H.
  {Connell}, J.~L. {Contreras}, M.~K. {Daniel}, M.~{Fornasa}, M.~{Gaug}, J.~F.
  {Glicenstein}, A.~{Gonz{\'a}lez-Mu{\~n}oz}, T.~{Hassan}, D.~{Horns},
  A.~{Jacholkowska}, C.~{Jahn}, R.~{Mazini}, N.~{Mirabal}, A.~{Moralejo},
  E.~{Moulin}, D.~{Nieto}, J.~{Ripken}, H.~{Sandaker}, U.~{Schwanke},
  G.~{Spengler}, A.~{Stamerra}, A.~{Viana}, H.-S. {Zechlin}, S.~{Zimmer}, and
  {CTA Consortium}, {\it {Dark matter and fundamental physics with the
  Cherenkov Telescope Array}},  {\em Astroparticle Physics} {\bf 43} (Mar.,
  2013) 189--214, [\href{http://arxiv.org/abs/1208.5356}{{\tt
  arXiv:1208.5356}}].

\bibitem{2012arXiv1206.6288A}
E.~{Aprile} and {XENON1T collaboration}, {\it {The XENON1T Dark Matter Search
  Experiment}},  {\em ArXiv e-prints} (June, 2012)
  [\href{http://arxiv.org/abs/1206.6288}{{\tt arXiv:1206.6288}}].

\end{thebibliography}\endgroup

\end{document}